\documentclass[aps,groupedaddress,superscriptaddress,showpacs,twocolumn,amsmath,amssymb,prl]{revtex4}
\usepackage{dcolumn}
\usepackage{bm}
\usepackage[dvips]{graphicx}
\usepackage{wrapfig,subfigure}
\usepackage{amssymb}
\usepackage{color}
\usepackage{ulem}
\usepackage{soul}

\newcommand{\bb}[1]{{\boldsymbol{#1}}} 

\begin{document}
\title{Flux $1/f^{\alpha}$ noise in 2D Heisenberg spin glasses: effects of weak anisotropic interactions}
\author{Juan Atalaya}
\affiliation{Institut f\"ur Theoretische Festk\"orperphysik, Karlsruhe Institute of Technology, 76131 Karlsruhe, Germany}
\affiliation{DFG Center for Functional Nanostructures (CFN), Karlsruhe Institute of Technology, 76131 Karlsruhe, Germany}
\affiliation{Department of Physics and Astronomy, Michigan State University, East Lansing, Michigan 48824, USA}
\author{John Clarke}
\affiliation{Department of Physics, University of California, Berkeley, California 94720-7300, USA}
\author{Gerd Sch\"on}
\affiliation{Institut f\"ur Theoretische Festk\"orperphysik, Karlsruhe Institute of Technology, 76131 Karlsruhe, Germany}
\affiliation{DFG Center for Functional Nanostructures (CFN), Karlsruhe Institute of Technology, 76131 Karlsruhe, Germany}
\author{Alexander Shnirman}
\affiliation{Institut f\"ur Theorie der Kondensierten Materie, Karlsruhe Institute of Technology, 76131 Karlsruhe, Germany}
\affiliation{DFG Center for Functional Nanostructures (CFN), Karlsruhe Institute of Technology, 76131 Karlsruhe, Germany}

\begin{abstract}
We study the dynamics of a two-dimensional ensemble of randomly distributed classical Heisenberg spins with isotropic RKKY and weaker anisotropic dipole-dipole couplings. Such ensembles may give rise to the flux noise observed in SQUIDs with a $1/f^{\alpha}$ power spectrum with $\alpha\lesssim1$. We solve numerically the Landau-Lifshitz-Gilbert equations of motion in the dissipationless limit. We find that Ising type fluctuators, which arise from spin clustering close to a spin-glass critical behavior with $T_c=0$, give rise to $1/f^{\alpha}$ noise. Even weak anisotropic interactions lead to a crossover from the Heisenberg-type criticality to the much stronger Ising-type criticality. The temperature dependent exponent $\alpha(T)\lesssim1$ increases and approaches unity when the temperature is lowered. This mechanism acts in parallel to the spin diffusion mechanism. Whereas the latter is sensitive to the device geometry, the spin-clustering mechanism is largely geometry independent.   
\end{abstract}
\pacs{85.25.Dq, 75.10.Nr, 75.40.Mg, 74.40.De}

\maketitle
\section{Introduction}
Excess low-frequency flux noise is a ubiquitous phenomenon observed in superconducting quantum interference devices 
(SQUIDs)~\cite{Wellstood_JClarke1987,Sendelbach_McDermott2008,Sendelbach_McDermott2009,Drung2011,JClarke2013}, and  
flux~\cite{Yoshihara2006,Kakuyanagi2007,RHarris_DVAverin2008,SAnton_JClarke2012,FYan_WOliver2012} 
 and phase~\cite{RBialczak_JMartinis2007,DSank_JMartinis2012} qubits
down to arbitrarily low temperatures. Various experiments have shown that flux noise has almost sample independent properties: the power spectrum at low frequencies scales as $1/f^{\alpha}$ with $\alpha\lesssim1$ with an amplitude of order $1\mu\Phi_0$~Hz$^{-1/2}$ at $f=1$~Hz, where $\Phi_0\equiv h/2e$ is the flux quantum. 
Understanding the mechanism producing this flux noise and developing strategies to reduce it are crucial steps towards improving the performance of superconducting devices in quantum information processing~\cite{Makhlin2001,Paladino2013} and for many applications of SQUIDs in, \textit{e.g.}, medicine and metrology~\cite{JC_AB2004,Drung2011}. 

Recent experiments suggest that the observed flux noise is produced by the slow dynamics of paramagnetic spins localized at metal-substrate or metal-surface oxide interfaces
with areal spin density $\rho_{s}\sim0.5$~nm$^{-2}$~\cite{Koch2007,Sendelbach_McDermott2008,HBluhm_KMoler2009}. It has been proposed that these spins are due to metal-induced gap states localized at metal-insulator interfaces~\cite{SChoi_JClarke2009}.
In proximity to a metal the spins interact via the isotropic long-range RKKY interaction~\cite{RKKY3}. For the typical spin separation of order $r_{\rm typ}=1.5$~nm the interaction strength is $J_0=50$~mK~\cite{Faoro_Ioffe2008}, corresponding to a frequency $f\approx1$~GHz. (We use units where $\hbar=k_{\rm B}=1$.) However, for pairs with (minimal) separation $r_{\rm min}=0.3$~nm, corresponding to a typical atomic spacing, the interaction strength reaches much larger values, $J_{0\max}=6.25$~K.  The random distances lead to competing ferromagnetic and antiferromagnetic interactions. In addition, the spins are coupled via the dipole-dipole interaction which, for spins with magnetic moment $\mu_{\rm B}$, has strength $J_1=0.55$~mK and $J_{1\max}=70$~mK for typical and minimal separations, respectively. The dipolar coupling gives rise to an anisotropic interaction depending on the orientation of the spins with respect to their relative positions. This leads to qualitatively different behavior, which is the focus of this paper.

The dynamics of spin ensembles and their effects on the flux noise spectra have been studied previously:

\noindent
(1) Faoro and Ioffe~\cite{Faoro_Ioffe2008} considered a model with RKKY coupling and found that diffusion of the spin magnetization produces flux noise with a $1/f$-type power spectrum for frequencies $f\gtrsim f_l\sim\mathcal{D}W^{-2}$, where $\mathcal{D}$ is the spin diffusion coefficient, $W$ the SQUID loop width and $f_l$ a low frequency cut-off. Without further assumptions, the spectrum is temperature independent. To reconcile the model with experimental data, showing an exponent $\alpha(T)\lesssim1$ which grows as $T$ decreases below $\sim1$~K \cite{Wellstood_JClarke1987,Wellstood2011,JClarke2013}, it has been conjectured that the spin diffusion coefficient depends on temperature~\cite{Lanting2013}. Concerning this temperature dependence for Heisenberg spin glasses (HSGs) above the spin glass (SG) transition there are contradictory points of view~\cite{Hertz1979,Gotze84}.

\noindent
(2) It has been proposed~\cite{Ioffe2012} that pairs of atypically strongly coupled spins exhibit slow switching dynamics between triplet and singlet states, driven by the high-frequency noise from the remaining paramagnetic spins. The influence of spin clusters on magnetic noise has been analyzed~\cite{Glazov2014}. In particular, clusters of ferromagnetically coupled F$^+$ point defects in amorphous insulating materials (\textit{e.g.}, Al$_2$O$_3$) may lead to correlated flux and inductance noise with $1/f^{\alpha}$ spectra~\cite{Amrit2014}. 

\noindent 
(3) Monte Carlo simulations~\cite{Zchen_CYu2010} of a 2D short-range Ising spin glass (ISG) show that at low temperatures the total magnetization noise exhibits $1/f$-like properties even below the lower critical dimension $d_l^{\rm Ising}\approx2.5$ of the ISG phase transition~\cite{BrayMoore84,Fischer1991}. The physical origin of Ising spins, however, remained unexplained. 

\noindent
(4) Isotropic HSG models have a lower critical dimension $d_l^{\rm Heis}=3$ or $4$ for RKKY-type or short-range interactions, respectively~\cite{Kawamura2003,BrayMooreYoung86}. Thus, no spin-freezing occurs in these models in 2D at $T\ne0$. However, critical behavior is observed in numerical simulations at low temperatures when the SG correlation length, diverging as $\xi(T)\sim(T-T_c)^{-\nu}$ with 
critical temperature $T_c=0$, becomes larger than $r_{\rm typ}$. The critical exponents are $\nu\sim1$ for the 2D HSG~\cite{Kawamura2003} and $\nu\sim3$ for the 2D ISG~\cite{BrayMoore84}. Furthermore, it is known that even weak anisotropic couplings make the HSG exhibit properties of an ISG~\cite{Fischer1991}. 

\section{Model}
To resolve the questions left open by the models discussed above and their implications for the flux noise spectrum,
we performed numerical studies of the 
dynamics of a 2D random ensemble of classical Heisenberg spins. We assume the spin dynamics to be described by the Landau-Lifshitz-Gilbert equation~\cite{Brown1963,PhDepondt2009} 
\begin{equation}
\label{eq:LLG}
\dot{\bb{s}}_i (t) = -\bb{s}_i\times\big(-\partial_{\bb{s}_i}{H} + \bb{h}_{i}(t) - {\eta}\dot{\bb{s}}_i\big)\ .
\end{equation}  
Here we use units such that $|\bb{s}_i|=1$ and the damping parameter $\eta$ is dimensionless. The noise fields $\bb{h}_i(t)$ are assumed to be Gaussian distributed with correlators $\langle h_{i\nu}(t)h_{j\nu'}(t')\rangle=2T\,\eta\,\delta_{ij}\,\delta_{\nu\nu'}\,\delta(t-t')$ and $\nu=x,y,z$. The Hamiltonian is given by  
\begin{eqnarray}\label{eq:Hamiltonian}
H&=& \sum_{(ij)} J_{0ij} \, \bb{s}_{i}\cdot\bb{s}_{j} \nonumber\\
&-& \sum_{(ij)} \,J_{1ij}\,  \Big[\frac{3(\bb{s}_i\cdot\bb{r}_{ij})(\bb{s}_j\cdot\bb{r}_{ij})}{r_{ij}^2} - \bb{s}_i\cdot\bb{s}_j\Big]\ ,  
\end{eqnarray}
where $J_{0ij}$ are the isotropic RKKY couplings (with varying sign) and $J_{1ij}>0$ the anisotropic dipole-dipole couplings between 
spins $i$ and $j$, separated by $\bb{r}_{ij}\equiv\bb{r}_i-\bb{r}_j$. Both coupling strengths vary with distance  $\propto 1/r_{ij}^{3}$. As a result, for those spins in the random ensemble which are separated by distances smaller than $r_{\rm typ}$, the couplings are much stronger than the typical values $J_0$ and $J_1$. 

In our simulations, we actually assume the spins to be placed on a $N\times N$ square lattice with lattice constant $a=r_{\rm typ}$, and we include only nearest neighbor couplings.
To account for the variation of coupling strengths we set $J_{0ij}=\pm J_{0}(r_{\rm typ}/r_{ij})^3$ (with random signs) and $J_{1ij}=J_{1}(r_{\rm typ}/r_{ij})^3$, choosing the random distances $r_{ij}$ according to the distribution $P(r)=2\,r\,r_{\rm typ}^{-2}e^{-(r^2-r_{\min}^2)/r_{\rm typ}^2}\Theta(r-r_{\min})$. 
The exponential accounts for the decreasing probability for spins with separation much larger than $r_{\rm typ}$ to be nearest neighbors. In this work, we concentrate on dissipationless spin dynamics, $\eta=0$, with one exception: To introduce temperature, we start the simulations with nonzero $\eta$ and switch $\eta$ off only when the system is thermalized; see  Appendix A.  

From the spin trajectories, $\bb{s}_i(t)$, we obtain the discrete spatial Fourier components of the magnetization  $\bb{s}(\bb{k},t) = N^{-1}\cdot\sum_i\bb{s}_i(t)e^{i\bb{r}_i\cdot\bb{k}}$ and subsequently the power spectra $\mathcal{S}_{\nu}(\bb{k},\omega)=\langle|\mathcal{F}_{\omega}[s_{\nu}(\bb{k},t)-\langle s_{\nu}(\bb{k},t)\rangle]|^2\rangle$. Here 
$\mathcal{F}_{\omega}$ denotes the time-Fourier transform.
The spins on the surface of the SQUID lead to a total flux threading the SQUID loop~\cite{Anton2013,Faoro_Ioffe2008}  
\begin{equation}
{\Phi(t)} = \sum_{\bb{k},\nu} \mathcal{B}_\nu (\bb k){s}_\nu(\bb k,t)\ ,
\label{eq:fluxnoise}
\end{equation}
and their fluctuations lead to the flux noise power spectrum 
$\mathcal{S}_{\Phi}(\omega)\approx\sum_{\bb k,\nu}\mathcal{B}^2_\nu(\bb k)\mathcal{S}_{\nu}(\bb k,\omega)$.
The form factor $\mathcal{B}_\nu(\bb k)$ depends on the geometry. 
If we consider a planar, square-shaped SQUID with narrow lines of width $W$,
the flux noise contribution from one side of the square (along the $y$-direction with a narrow width in the $x$-direction) is given by~\cite{Faoro_Ioffe2008}
$\Phi_1(t)\approx\sum_{k_x}\mathcal{B}_x(k_x,k_y=0){s}_x(k_x,k_y=0,t)$
with $\mathcal{B}_x(k_x,k_y=0)\sim|k_x|^{-1/2}$~\cite{Lanting2013}. 
  For purely isotropic interactions, where the total magnetization is conserved, 
the spin dynamics should reduce to spin diffusion, \textit{i.e.} $\mathcal{S}_{\nu}(\bb k,\omega)\propto2\Gamma_k(\Gamma_k^2+\omega^2)^{-1}$, with relaxation rate $\Gamma_k=\mathcal{D}k^2$. In combination with the form factor given above spin diffusion leads to a power spectrum $\mathcal{S}_{\Phi}(\omega)\sim\omega^{-1}$ within the frequency range $\omega_{l}\le\omega\le\omega_{u}$, where $\omega_{l}\sim\mathcal{D}(2\pi/W)^{2}$ and the upper frequency cut-off $\omega_{u}\sim\mathcal{D}(2\pi/r_{\rm typ})^{2}$~\cite{Faoro_Ioffe2008}. 
  
 \begin{figure}[t!]
	\centering
\includegraphics[width=\linewidth,trim=0cm 0cm 0cm 0cm, clip=true]{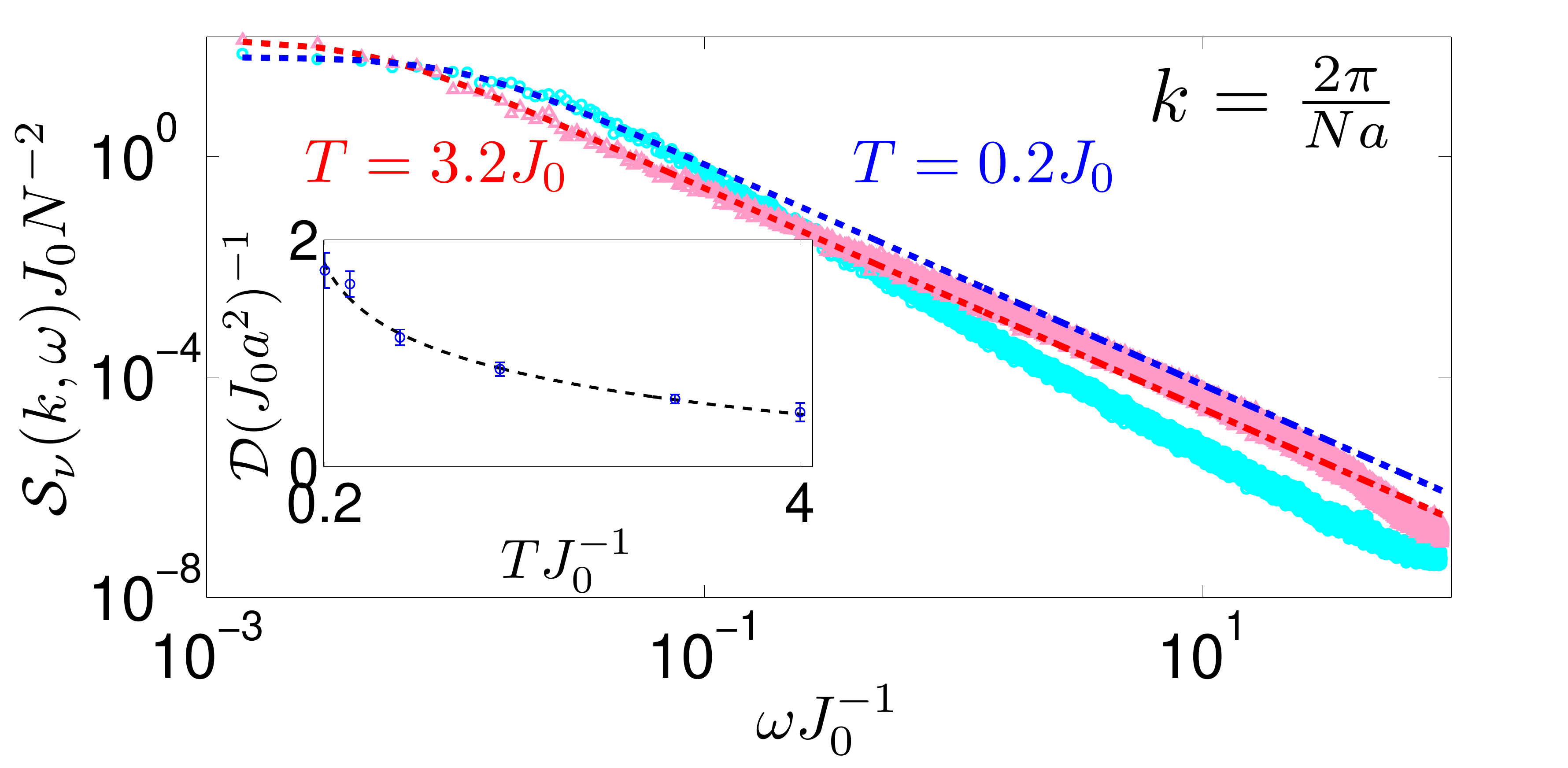}
\vspace{-6mm}
\caption{(Color online) Spin fluctuation power spectra $\mathcal{S}_{\nu}(k,\omega)$ for the case of purely isotropic interactions. The spectra are independent of the spin direction $\nu=x,y,z$. 
Temperatures are $T=0.2J_0$~(circles) and $3.2J_0$ (triangles). 
The system size is $N^2$ spins with $N=80$. Dashed lines are Lorentzian fits. 
Inset: $T$-dependence of the spin diffusion coefficient $\mathcal{D}(T)$. 
The dashed line is a logarithmic fit. 
\label{fig:fig1}}
\end{figure} 

\section{Isotropic Interactions}
To analyze the conditions for spin diffusion and a potential temperature-dependence of $\mathcal{D}$ we first perform simulations within our model for a 2D HSG  with purely isotropic interactions ($J_{1ij}=0$). Figure~\ref{fig:fig1} depicts the power spectra $\mathcal{S}_{\nu}(k,\omega)$ for 
two temperatures. We note that the spectra fit well to Lorentzians of the form $2C_k\Gamma_k(\Gamma_k^2+\omega^2)^{-1}$. The fit is perfect at high temperatures, while at lower values, $T\ll J_0$, it is valid only up to a cut-off frequency, which decreases as $T$ is lowered.
The  parameter $C_{k}$  is proportional to the static susceptibility for wavenumber $k$ and approximately independent of $k$. The relaxation rates are well described by $\Gamma_k=\mathcal{D}k^2$. At high temperatures $T\gtrsim J_0$ the diffusion coefficient $\cal{D}$ is constant, while at lower temperatures it is weakly temperature-dependent (see inset of Fig.~\ref{fig:fig1}). The dependence appears to be logarithmic, consistent with the results of Ref.~\cite{Hertz1979}, $\mathcal{D}\propto(T-T_c)^{2/d-1}$ ($2<d<4$), extrapolated to $d=2$, where $T_c=0$~\cite{Kawamura2003}. 

\begin{figure}[t!]
	\centering
\includegraphics[width=\linewidth,trim=0cm 0cm 0cm 0cm, clip=true]{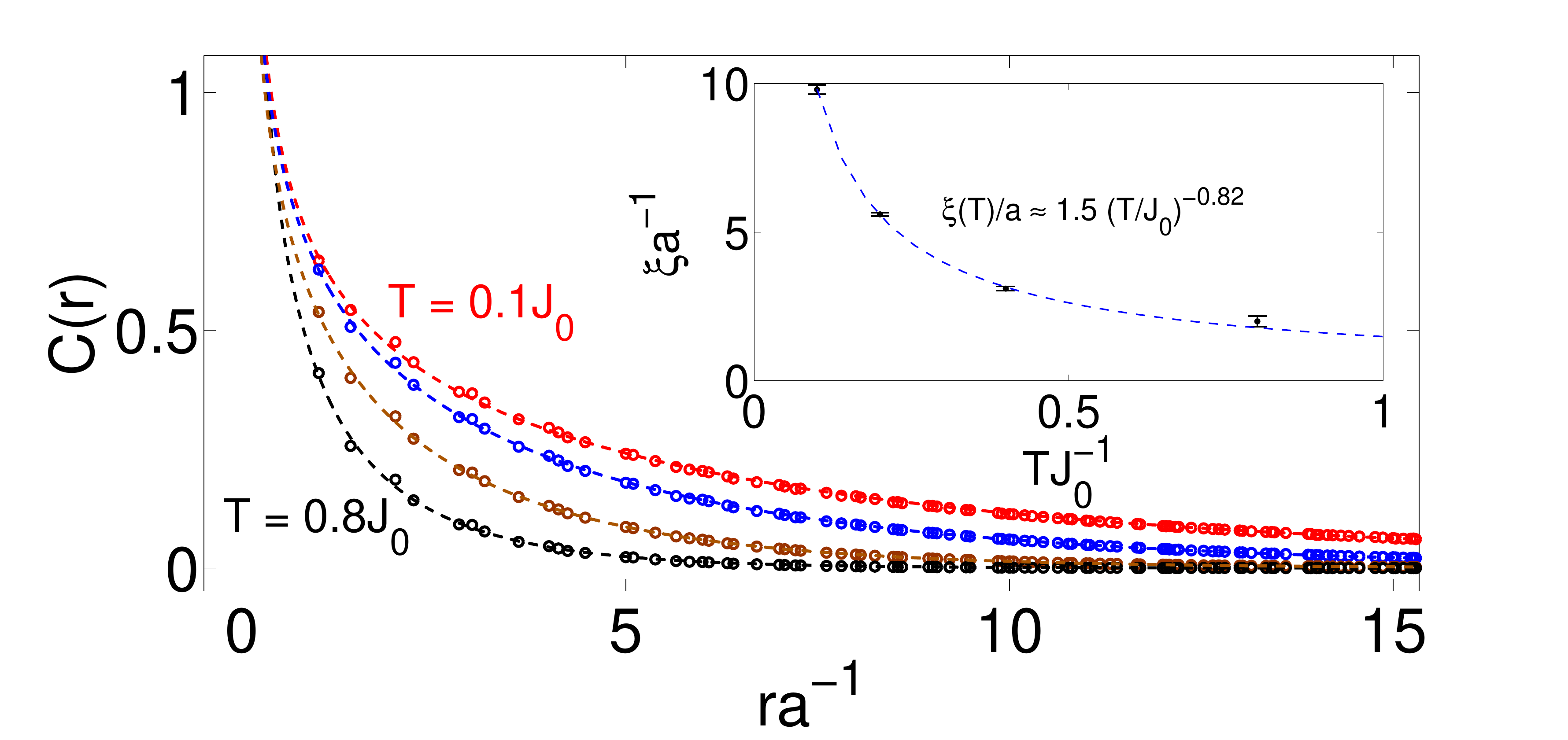}
\vspace{-6mm}
\caption{The spin glass correlation function, $C(r)$, for temperatures  $T=0.1J_0$, $0.2J_0$, $0.4J_0$ and $0.8J_0$. The dashed lines are fits to $C(r) \propto e^{-r/\xi(T)}/r^b$. The inset depicts the temperature dependence of the spin glass correlation length $\xi(T)$. The system size is $N^2$ spins with $N=80$. The number of different realizations of the random coupling is four. 
\label{fig:fig1B}}
\end{figure}

Our data also show evidence of critical behavior. Figure~\ref{fig:fig1B} shows the spin glass (static) correlation function, $C(r)\equiv N^{-2}\sum_{ij}\big \langle [\langle {\bf S}_i\cdot {\bf S}_j\rangle - \langle{\bf S}_i\rangle\cdot\langle{\bf S}_j\rangle]^2\big\rangle_{\rm c}\delta(r-r_{ij})$ where $\big\langle\cdot\big\rangle_{\rm c}$ denotes the averaging over different realizations of the random coupling, for different temperatures. At low temperatures the correlation function can be fitted to $C(r)=Ae^{-r/\xi(T)}/(r/a)^b$, where $\xi(T)$, $A\approx0.7$ and $b\approx0.3$ are fitting parameters. The correlation length $\xi(T)$ diverges at low temperatures as $\xi(T)\approx 1.5a\big[(T-T_c)J_0^{-1}\big]^{-\nu}$ with critical exponent $\nu\approx0.8$ and~$T_c=0$. 

\section{Anisotropic Interactions}
We turn next to spin dynamics in the presence of weak dipolar couplings $J_{1ij} \ll |J_{0ij}|$. 
Such couplings break the rotational symmetry, \textit{i.e.}, the results depend on the spin direction,  $\nu=x,y,z$, and the total magnetization is no longer conserved.
As with the RKKY coupling, $J_{1ij}$ depends strongly on the separation. Our simulations reveal
that the maximum anisotropic coupling $J_{1\max}$ determines the crossover temperature where qualitative changes in the spin dynamics are observed.

At high temperatures $T \gtrsim J_{1\max}$ the noise spectra with small $k$ still have a Lorentzian shape with relaxation rates $\Gamma_{\nu}(k)=\Gamma_{\nu,0}+\mathcal{D}k^2$, where $\Gamma_{\nu,0}=\mathcal{A}_{\nu}\langle J_{1ij}^2\rangle J_0^{-1}$ and the numerical coefficients are $\mathcal{A}_{z}\approx2\mathcal{A}_{x,y}\approx2$ and $\langle J_{1ij}^2\rangle\approx J_1^2r_{\rm typ}^4/2r_{\rm min}^{4}$. The dynamics in this temperature regime is of the Heisenberg type, \textit{i.e.}, the spins explore the entire Bloch sphere, except that the out-of-plane magnetization fluctuations relax faster than the in-plane ones. As a result of the relaxation, the cut-off $f_l$ for the $1/f$-like flux noise spectrum,
which arises from the specific geometrical form factor~\cite{Faoro_Ioffe2008},  
is now given by $f_l\sim\max\{\mathcal{D}(2\pi/W)^2,\Gamma_{\nu,0}\}/2\pi$. As we argue below this rules out the diffusion mechanism as the origin of the observed $1/f$ noise at low frequencies. 		

\begin{figure}[t!]
\centering
\begin{tabular}{c}
\includegraphics[width=1.4\linewidth,trim=0cm 0cm 0cm 0cm, clip=true]{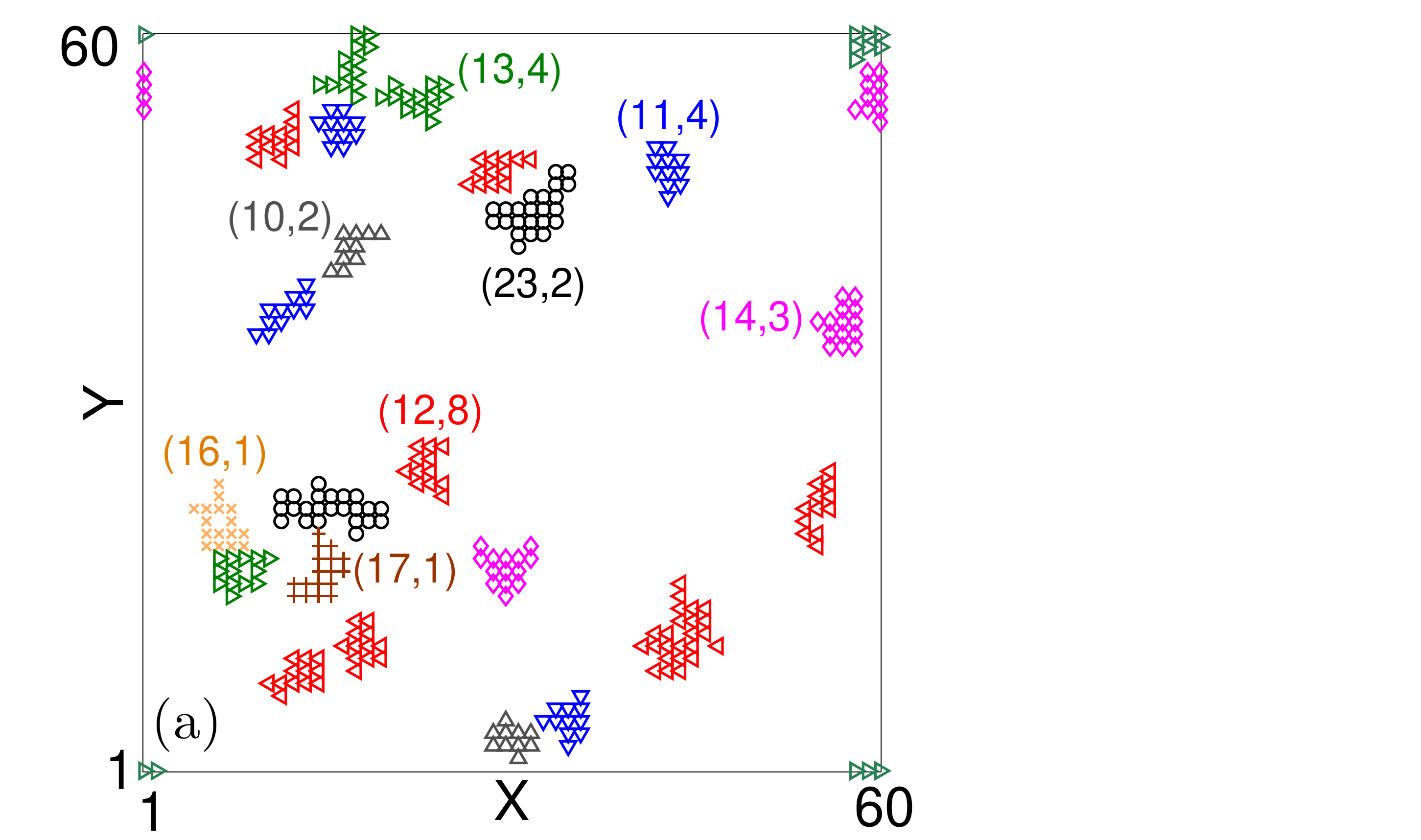} 
\end{tabular}
\begin{tabular}{c}
\includegraphics[width=\linewidth,trim=0cm 0cm 0cm 0cm, clip=true]{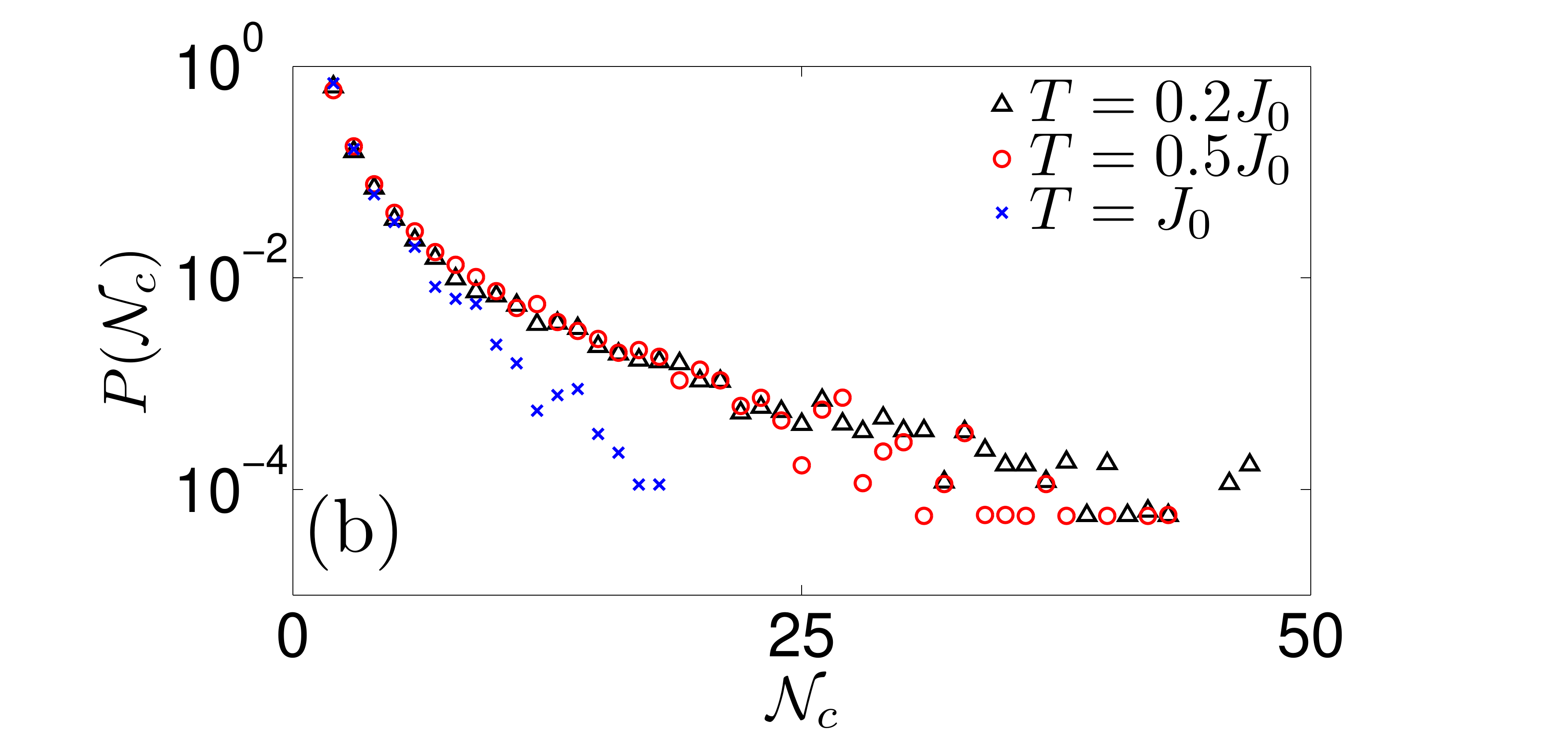}
\end{tabular}
\caption{(Color online) Distribution of clusters defined as groups of spins for which the magnitude of the mutual correlator, $c_{\nu\,ij}\equiv\langle s_{\nu i}s_{\nu j}\rangle-\langle s_{\nu i}\rangle\langle s_{\nu j}\rangle$, exceeds $0.95$. (a) Large clusters ($\nu=x$) of sizes between 10 ($\triangle)$ and 23 ($\circ$) are depicted for a given realization of the random couplings at temperature $T=0.5J_0$. The labels $(\mathcal{N}_c,n)$ indicate the number of clusters, $n$, with size $\mathcal{N}_c$. The system size is $60\times60$ spins. (b) Probability distribution, $P(\mathcal{N}_c)$, for the sizes of clusters (with magnetization parallel and perpendicular to the plane) at different temperatures.
\label{fig:fig2}}
\vspace{-2mm}
\end{figure}  

For lower temperatures $T\lesssim J_{1\max}$ the dynamics changes from Heisenberg to Ising type. At the crossover $T\approx J_{1\max}$ a Griffiths SG phase develops~\cite{Griffiths69,Palmer1985} and the first pairs of closely spaced --and thus strongly coupled-- spins begin to form. Each pair behaves as a two-state fluctuator with the two energy minima corresponding to both spins pointing in the same direction, but either parallel or antiparallel to the vector connecting their positions. 
In the absence of RKKY interactions the energy barrier between these two minima would be $\Delta U=J_{1ij}$, but in the presence of strong RKKY coupling it is enhanced, $\Delta U=3J_{1ij}$,
 provided the coupling happens to be ferromagnetic. 
Strongly coupled pairs with antiferromagnetic RKKY interaction tend to form singlets. They
do not contribute to the flux noise, but they may  contribute to the magnetic 
susceptibility noise due to rare thermal jumps to the triplet state~\cite{Ioffe2012}.

\begin{figure}[t!]
\centering
\includegraphics[width=\linewidth,trim=0cm 0cm 0cm 0cm, clip=true]{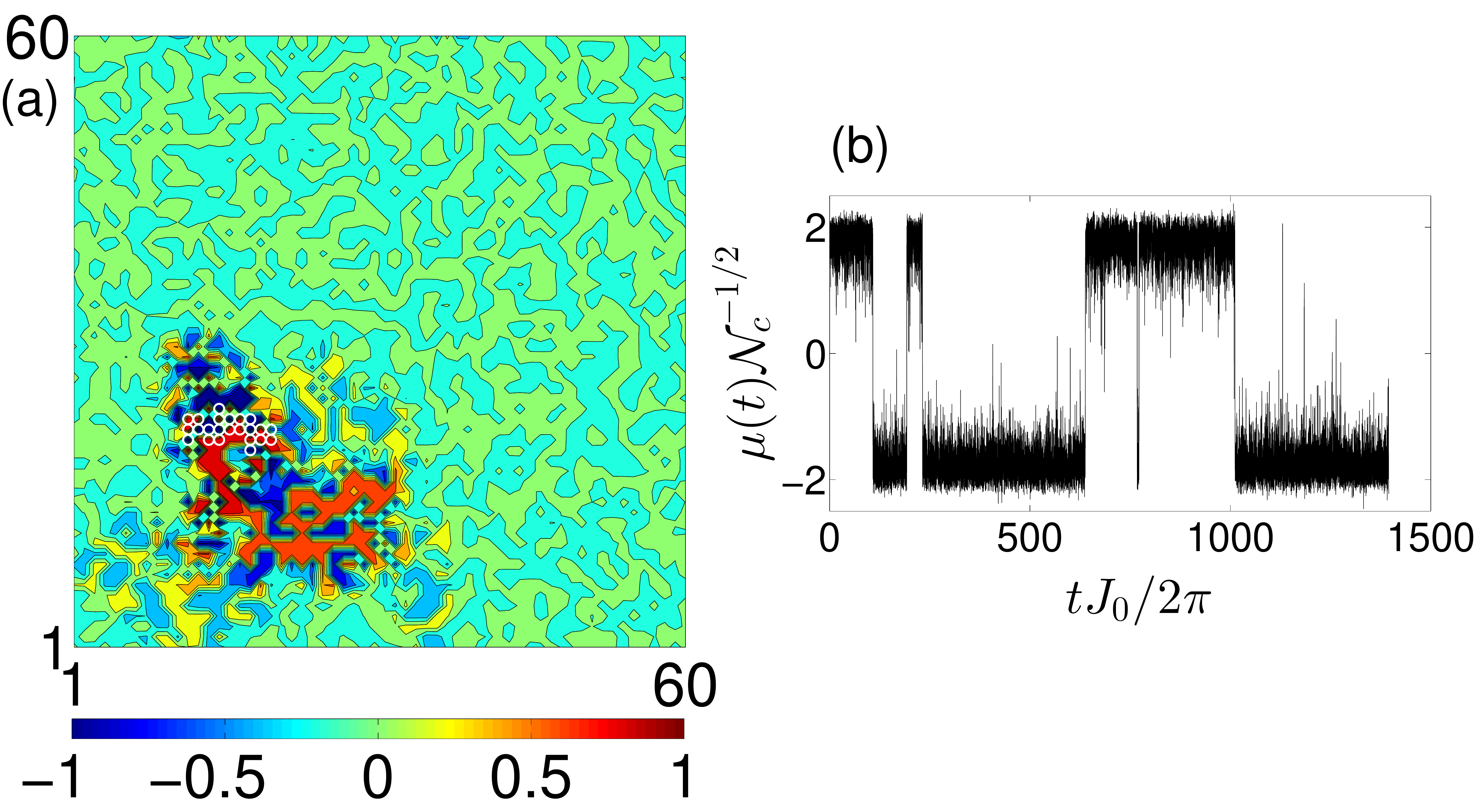}
\caption{Properties of an Ising-like cluster. (a) Spatial distribution of the correlator $c_{x\,ij}$ between one spin ($i$) belonging to a cluster of size $\mathcal{N}_c=23$ (white circles) and the other spins ($j$). Configuration and parameters as in Fig.~\ref{fig:fig2}(a). Red and blue colors denote ferromagnetic and antiferromagnetic correlations, respectively. (b) Time dependence of the magnetization, $\mu(t)$, of the cluster shown in (a).
\label{fig:fig3}}
\vspace{-5mm}
\end{figure}

The distribution of energy barriers, related to the range of spin separations $r_{ij}$ discussed above, leads to a $1/f$-like flux noise spectrum up to temperatures of order $T\lesssim  J_{1\max}$. In addition, as illustrated in Fig.~\ref{fig:fig2} and Fig.~\ref{fig:fig3}(a), at lower $T$ more complicated cluster configurations emerge with a temperature-dependent distribution of sizes $\mathcal{N}_c$. The larger clusters turn out to be mostly random (glassy) with magnetization of order $\mu_c\sim\sqrt{\mathcal{N}_c}$. 
The magnetization of the clusters, $\mu(t)$, switches in Ising-type fashion between the values $\pm\mu_c$, as illustrated in Fig.~\ref{fig:fig3}(b). The larger the cluster the slower is the switching. 
For  large clusters both ferromagnetic  and 
antiferromagnetic RKKY couplings raise the energy barriers. 
The dynamics of the clusters are predominantly driven by the surrounding bath of paramagnetic spins. In comparison, the coupling between clusters is negligible in the dilute system considered. In the opposite limit of a non-dilute system, the cluster dynamics are similar to those of the (coarse-grained) spin-glass models studied previously~\cite{Zchen_CYu2010}. 

As a result of the switching dynamics the power spectra for the total magnetization, as well as the Fourier modes with non-zero $k$, exhibit a $1/f^{\alpha}$ dependence in a frequency range $f_l\lesssim f\lesssim f_u$, as shown in Fig.~\ref{fig:fig4}.  The upper cut-off frequency
 scales with the strength of the anisotropic coupling $f_u=A J_{1\max}/2\pi$, where $A\sim1$ (but $<1$).  
Because of limited simulation time we can not precisely determine the lower cut-off 
frequency $f_l$. 
Figure~\ref{fig:fig4} also shows that the exponent $\alpha(T)<1$ tends to increase as $T$ is lowered. At higher frequencies $f_u\lesssim f\lesssim J_{0\max}/2\pi$, the spectral functions corresponding to small wave numbers $k$ decay roughly as $f^{-3}$, whereas those with larger $k$ decay as $f^{-2}$. For even higher frequencies $f\gtrsim J_{0\max}/2\pi$, the spectra decay  rapidly depending on the distribution of the coupling strengths. Since finite-$k$ Fourier modes also exhibit $1/f^{\alpha}$-like spectra, the flux noise spectrum arising from the magnetization switching dynamics of the clusters is not significantly affected by the SQUID geometry and form factor. 

At still lower temperatures $T\lesssim J_1$ the system shows a tendency towards spin-glass freezing. 
Since the lower critical dimension of an ISG is higher than 2, one expects $T_c=0$, and the observed freezing is a consequence of finite size, with $\xi(T)\gtrsim W$, and limited simulation time.  	

\section{Comparison with experiments}
\begin{figure}[t!]
	\centering
\includegraphics[width=\linewidth,trim=0cm 0cm 2cm 0cm, clip=true]{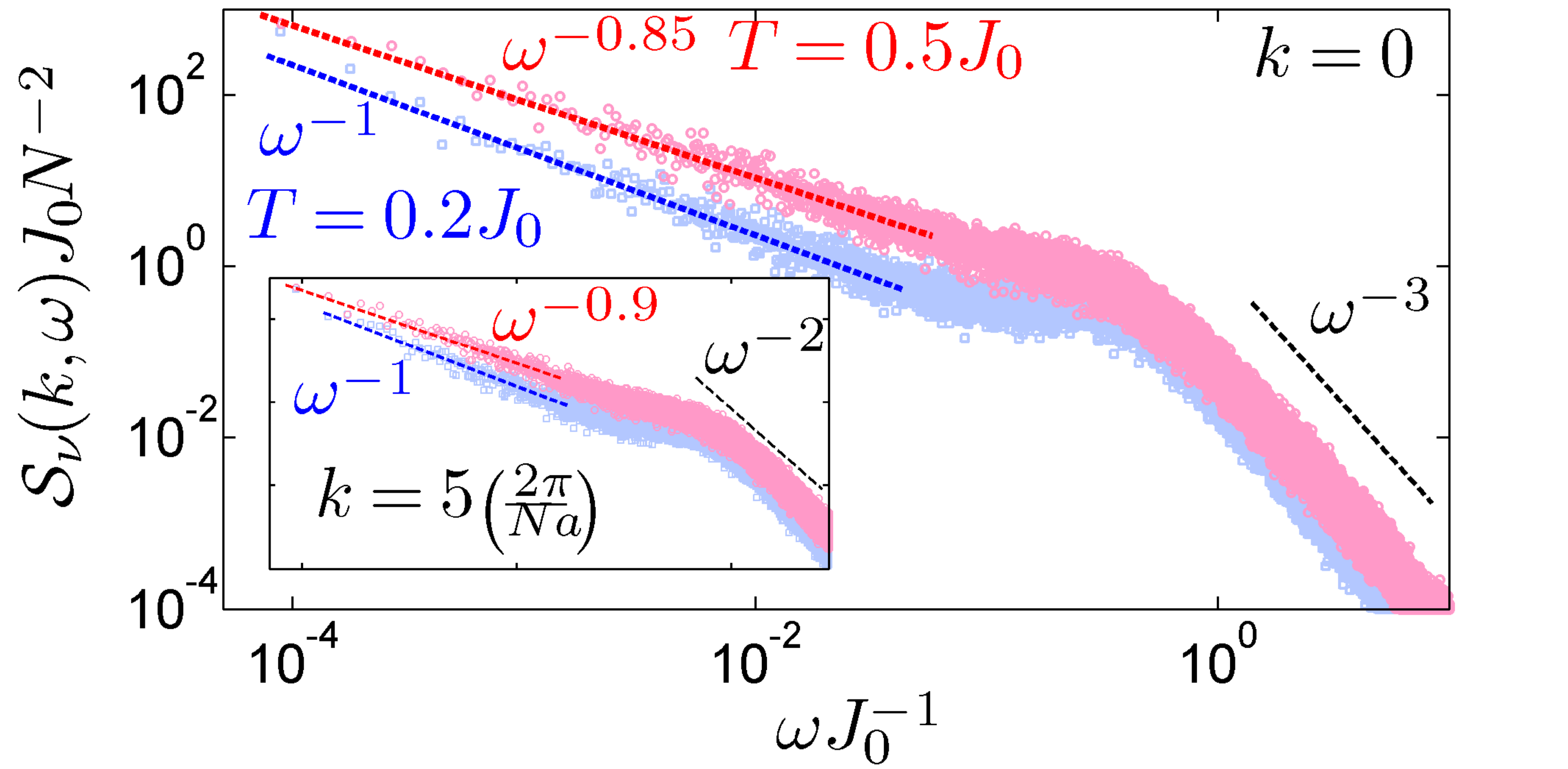}
\vspace{-6mm}
\caption{Power spectra ${\cal S}_{\nu}(k,\omega)$ of the in-plane ($\nu=x,y$) magnetization fluctuations, 
for different wave numbers and temperatures for a system with $N^2$  spins ($N=60$). 			
\label{fig:fig4}}
\vspace{-2mm}
\end{figure}

As shown in Fig.~\ref{fig:fig4} (with $J_0=2\pi\times1$~GHz) we find a $1/f^{\alpha}$ flux noise spectrum  in the frequency range between $f_l<100$~kHz and $f_u\sim200$~MHz. While the upper cut-off $f_u$ is readily accessible from our numerical analysis and roughly agrees with experiments~\cite{Bylander2011,Siddiqi2012}, the lower cut-off  $f_l$ cannot be  determined because of the limitations of simulation time. We conjecture  that the appearance of large clusters, larger than those we can simulate, would lead to $1/f$ noise down to much lower frequencies. 

To compare with the observed magnitude of the flux noise power spectrum $\mathcal{S}_{\Phi}(\omega)$ for SQUIDs we have to extrapolate our data. Our simulation is restricted to a square of size $90\times90$~nm$^2$ (based on $a=1.5$~nm), while the line width  $W$ and outer dimension of the SQUID loop $L\approx10W$ are typically much larger. Assuming that fluctuations from different squares contribute independently, we 
find $\mathcal{S}_{\Phi}(\omega)=\rho_s\bar{g}\,(L/W)(\mu_B\mu_0)^2N^{-2}\mathcal{S}_{x,y}(k=0,\omega)$, where $\bar{g}\approx3.5$ is the squared form factor~\cite{Faoro_Ioffe2008} averaged over the SQUID line width.
From Fig.~\ref{fig:fig4} we find $\mathcal{S}_{\Phi}(f=100~\textrm{kHz})\approx2\times10^{-5}(\mu\Phi_0)^2$~Hz$^{-1}$ for $T=0.2J_0$. Extrapolating this result down to $f=1$ Hz (assuming an exponent $\alpha=1$) we find $\mathcal{S}_{\Phi}(f=1~\textrm{Hz})\approx2\,(\mu\Phi_0)^2$~Hz$^{-1}$, which is the same order of magnitude as the measured flux noise power spectrum. 

The fact that the cluster magnetic moment scales as  $\mu_c\propto\sqrt{\mathcal{N}_c}$ implies that the clusters are glassy (as opposed to ferromagnetic or antiferromagnetic). This has important implications. First, the mean square flux noise is independent of the cluster size only for glassy clusters~\cite{JClarke2013}. This is not the case for ferromagnetic or antiferromagnetic clusters. Second, the observed Curie-law paramagnetism~\cite{Sendelbach_McDermott2008} implies a classical unsaturated behavior, that is $\mu_cB<k_BT$, where $B$ is the ambient magnetic field. For $B=10$~mT and $T=50$~mK, we find $\mu_c\lesssim7\mu_B$ or $\mathcal{N}_c\lesssim49$.  In our simulations virtually all the clusters satisfy this restriction. For larger, experimentally relevant systems beyond the range of our simulations we expect larger, slower clusters to appear. These would be responsible for the $1/f^\alpha$ noise at low frequencies down to $10^{-4}$~Hz. Our estimates show, however, that the contribution to the susceptibility of these large clusters is masked by the much higher number of smaller clusters.   

Finally, we mention that our simulations provide indications of spin-glass freezing at low temperatures $T\lesssim J_1$. Our parameters translate to a system size $W\sim90$~nm, simulation time {$t_{\rm sim}=10^4(J_0/2\pi)^{-1}\approx10$~$\mu$s and freezing temperatures below $T\le1$~mK}. This is too low to be observed in experiments. On the other hand, spin freezing has been observed in SQUIDs~\cite{Sendelbach_McDermott2008} at $T\approx30$~mK, which may suggest an anisotropic coupling much stronger than the value assumed in this paper.     
 
\section{Conclusions}

In summary, in order to study the dynamics of 2D Heisenberg spin glasses we have performed numerical simulations of the Landau-Lifshitz-Gilbert equations in the dissipationless limit. 
For purely isotropic interactions the spin-diffusion description for the long-wavelength modes is valid at all temperatures as long as the system size exceeds the HSG correlation length, $W\gg\xi(T)\propto T^{-\nu}$ with $\nu\approx0.8$. 
At low temperatures, $T\lesssim J_0$, the diffusion coefficient $\mathcal{D}$ has a weak (logarithmic) temperature dependence consistent with Ref.~\cite{Hertz1979}. 

In the presence of weak dipole-dipole coupling at high temperatures, $J_{1ij}\lesssim T$, we find that the long-wavelength modes have relaxation rates  
$\Gamma_{\nu}(k)=\Gamma_{\nu,0}+\mathcal{D}k^2$, where $\Gamma_{\nu,0}=\mathcal{A}_\nu\langle J_{1,ij}^2\rangle/J_0$  accounts for the relaxation of the total magnetization along the $\nu$-direction ($\nu=x,y,z$) with $\mathcal{A}_{\nu}=\mathcal{O}(1)$. 
As a result, the low-frequency cut-off $f_l$ for the $1/f$-like flux noise spectrum in samples of width $W$
is given by $f_l\sim\max\{\mathcal{D}(2\pi/W)^2,\Gamma_{\nu,0}\}/2\pi$. For the parameters specified above with $\mathcal{D}\approx 2\times10^{-8}$ m$^2$s$^{-1}$ (see inset of Fig.~\ref{fig:fig1}) and $W \ge 1\mu$m we find $f_l\sim\Gamma_{\nu,0}/2\pi\sim50$~MHz. Thus, we conclude that, in the presence of anisotropic dipole-dipole coupling, the mechanism of Ref.~\cite{Faoro_Ioffe2008} does not explain the observed low-frequency flux noise. 

Our central result is that, at temperatures lower than the maximum anisotropic coupling strength, $J_{1\max}$, spin clusters develop with Ising-type dynamics of magnetization switching.  The cluster size  and the effective barriers against switching increase as the temperature is lowered. The resulting range of relaxation rates leads to a $1/f^{\alpha(T)}$ spectrum with an exponent $\alpha(T)\lesssim1$ which increases as temperature is lowered. The upper bound for the anisotropic coupling may be as large as $J_{1\max}=70$~mK, so that $1/f$ noise would start to appear at temperatures below roughly this value. The observation of $1/f$ noise up to temperatures of at least $4.2$~K suggests an even stronger anisotropic interaction. 
Spin-orbit coupling could produce such an anisotropy and also introduce damping. 
Both are strongly material dependent.
In contrast, the qualitative features of the mechanism analyzed in our paper are rather universal and independent of the type of anisotropy. 

\section{Acknowledgements}
We acknowledge stimulating discussions with K. Moler, H. Bluhm, R. McDermott, T. Vojta, S.M. Anton, J.S. Birenbaum, S.R. O'Kelley, and further members of our IARPA-funded collaboration. This research was
funded by the Intelligence Advanced Research Projects Activity (IARPA) through the US Army Research Office, the German Science Foundation (DFG), and the German-Israeli Foundation (GIF). 

\section{Appendix A}

Here, we specify the parameters used in our numerical simulations. The time step is $\Delta t=(512J_{\rm max})^{-1}$, where $J_{\max}$ is the magnitude of the largest coupling in the problem. 
We concentrate on dissipationless spin dynamics. In order to initialize the system with an energy corresponding to a given temperature, however, we begin the simulations with random initial conditions and dimensionless damping parameter ${\eta}=0.1$. We start with a high temperature $T=5J_{\max}$, and gradually reduce $T$ to the value of interest. Then we switch off the dissipation. This thermalization procedure occupies $10^8$ time steps. We assume periodic boundary conditions in all calculations. 

In the case of purely isotropic coupling, the magnetization fluctuations achieve ergodicity within the simulation time for the considered system size of $N^2$ spins with $N=80$. This means that the simulation time is larger than the largest relaxation time in the problem $\sim\Gamma_{k_{\rm min}}^{-1}$, where $k_{\rm min}=2\pi(Na)^{-1}$ is the smallest wavenumber. For a given realization of the random coupling and wavenumber $k$, the available data time series is divided into four segments of equal size and the spectral function for the given $k$ is calculated by averaging the spectra obtained in each segment. The spectral functions $\mathcal{S}_{\nu}(k,\omega)$, shown in Fig.~\ref{fig:fig1}, additionally involve an  averaging over four different realizations of the random coupling. We determine the relaxation rate $\Gamma_k$ of the diffusive mode with wavenumber $k$ by fitting  $\mathcal{S}_{\nu}(k,\omega)$  for $\nu=\{x,y,z\}$ with Lorentzians of the form $2C_{k}\Gamma_k/(\Gamma_k^2+\omega^2)$. We estimate the numerical errors in $\Gamma_k$ to be below 10\%. 

In the presence of anisotropic coupling and at low temperatures $T\ll J_{1\max}$, on the other hand, ergodicity is not achieved within the simulation time, $t_{\rm sim}$. This is because of the presence of large clusters which do not switch or switch only a few times within the simulation time. Thus, it is not possible to resolve the contribution of these clusters to the total magnetization noise power spectrum at the smallest accessible frequency $f_{\min}=t_{\rm sim}^{-1}$. For $f\gtrsim f_{\min}$, the power spectra scale as $\mathcal{S}_{\nu}(k,\omega)\sim \omega^{-\alpha}$ ($\alpha\lesssim1$). The spectra are calculated by averaging over eight different realizations of the random coupling. The calculations including anisotropic interaction are performed mostly for systems of smaller size, namely $60\times60$ spins. This size is still much larger than the size of the clusters, $\mathcal{N}_c$, found at low temperatures [$\mathcal{N}_c\lesssim 50$, see Fig.~\ref{fig:fig2}(b)], allowing us to investigate the low-frequency part of the spectrum within a reasonable computation time.

\end{document}